\documentclass[conference]{IEEEtran}
\usepackage[T1]{fontenc}
\IEEEoverridecommandlockouts
\usepackage{cite}
\usepackage{amsmath,amssymb,amsfonts}
\usepackage{algorithmic}
\usepackage{graphicx}
\usepackage{xcolor}
\usepackage{array}
\usepackage{multirow}
\usepackage{hyperref}
\usepackage[capitalize]{cleveref}
\usepackage{graphicx}
\usepackage{subcaption}
\usepackage{pifont}

\def\BibTeX{{\rm B\kern-.05em{\sc i\kern-.025em b}\kern-.08em
    T\kern-.1667em\lower.7ex\hbox{E}\kern-.125emX}}
\begin{document}

\title{Proof-of-Useful-Work Blockchain for Trustworthy Biomedical Hyperdimensional Computing}

\author{
\IEEEauthorblockN{
Jinghao Wen\textsuperscript{1}, Dongning Ma\textsuperscript{1}, Sizhe Zhang\textsuperscript{1}, Hasshi Sudler\textsuperscript{2}, Xun Jiao\textsuperscript{1}
}
\IEEEauthorblockA{
\textsuperscript{1}Department of Electrical and Computer Engineering, Villanova University, Villanova, USA\\
\textsuperscript{2}Internet Think Tank, Inc., Woodland Hills, USA\\
Email: \{jwen01, dma2, szhang6, xun.jiao\}@villanova.edu ~~~hasshi.sudler@inttk.org
}
}

\maketitle
\newcommand{\model}{\textbf{HDCoin}}
\begin{abstract}
Hyperdimensional Computing (HDC) is a promising bio-inspired learning paradigm for its advantage of balancing performance and efficiency and has been increasingly applied to the bio-medical domain. In bio-medical applications, trustworthiness such as replicability and verifiability of the trained learning models is crucial. In this work, we introduce \model, the first proof-of-useful-work blockchain framework for HDC. With \model, we transform the conventional energy-wasteful mining process into a competitive process for developing high-accuracy, trustworthy and verifiable hyperdimensional models. We explore four diverse biomedical datasets, and conduct an extensive design-space exploration of key HDC hyperparameters of blockchain miners such as dimensionality, learning rate, and retraining iterations for model performance, adaptive mining difficulty and fairness on proof-of-useful-work.

\end{abstract}


\section{Introduction}
\label{sec:intro}
Hyperdimensional Computing (HDC) is a bio-inspired machine learning scheme that leverages abstract patterns in the form of high-dimensional numerical vectors for learning tasks~\cite{kanerva2009hyperdimensional}. Recently, it has gained increasing attention for its advantage of efficiency across multiple applications, including biomedical applications such as human activity recognition~\cite{kim2018efficient, morris2019comphd}, seizure detection~\cite{burrello2018one, ge2021seizure, ge2022applicability} and drug discovery~\cite{ma2022molehd, yang2022automated}. A key challenge of developing learning models fpr biomedical applications is to ensure the trustworthiness of the trained models such as to train replicable models, and able to verify the models for their authenticity. 

Blockchain technology, with its decentralized and immutable nature, offers an ideal solution for guaranteeing the trustworthiness of learning models~\cite{rahimi2016robust}. Specifically, the concept of Proof-of-Useful-Work (PoUW) introduces a paradigm where the effort typically wasted in conventional cryptocurrency mining processes is repurposed to train high-accuracy models. While there are several related efforts attempting to implement PoUW for neural networks~\cite{liu2021proof}, there is no previous effort to implement it for HDC. In addition, HDC also differs from conventional neural networks in many aspects such as model architecture as well as training and inference process~\cite{ma2024hyperdimensional}.  

With such motivation, we introduce \model, the first proof-of-useful-work blockchain framework designed specifically for training hyperdimensional models for biomedical applications. Our main contributions are:

\begin{itemize}
    \item We propose \model. Models trained using \model ~are verified by all the miners on the blockchain network, and can be verified by the public once the information is added as another block onto the chain.
    \item Based on the unique characteristics of HDC, we proposed tailored algorithm for evaluation.
    We evaluate \model ~using four biomedical datasets and explore the impact of HDC-specific hyperparameters such as retraining rates and retraining iterations on model performance. In general, we observe that a lower retraining rate and more retraining iterations can relatively improve the performance of the model.
    \item We empirically explore the adaptive mining difficulty and the fairness of proof-of-use-work in \model. Results show that \model ~adequately supports adaptive mining difficulty by selecting different minimum configuration of the model to train, and can generally encourage a fair competition that miners who invests more training effort will have a higher chance of winning during the mining process.
\end{itemize}

\section{Related Works}
Recently, HDC has demonstrated strong potential across a variety of biomedical applications~\cite{amrouch2022brain}. HDC has found itself capable of ExG tasks for its unique encoding schemes, such as seizure detection, gesture classification and emotion recognition~\cite{rahimi2018efficient, moin2018emg, chang2019hyperdimensional}. HDC has also been applied to bioinformatics applications such as DNA pattern matching, drug discovery and genome sequencing for its efficiency and easier architectural support for acceleration~\cite{kim2020geniehd, ma2022molehd, zou2022biohd}.

As models for bioinformatics and biomedical applications are often applied to life-critical tasks, it is critical to ensure the reproducibility and verifiability of them, which naturally motivates us to consider blockchains. In blockchains, the concept of PoUW is a more beneficial alternative to PoW as it redirects solving hash problems into something more ``useful''~\cite{ball2017proofs}. Early PoUW methods targeted domains like scientific computing, protein folding, and SAT solving \cite{todorovic2022proof, fitzi2022ofelimos}, aiming to replace wasteful hashing with socially valuable computation. Recent efforts shift PoUW toward machine learning, particularly neural network training~\cite{liu2021proof, li2023proof}. As HDC is an emerging learning paradigm, no prior work has explored HDC with the concept of PoUW. \model~introduces the first framework to integrate HDC model training into blockchain mining, offering verifiability and adaptability tailored to biomedical learning.

\begin{figure}
\centering
    \includegraphics[keepaspectratio,width=1\columnwidth]{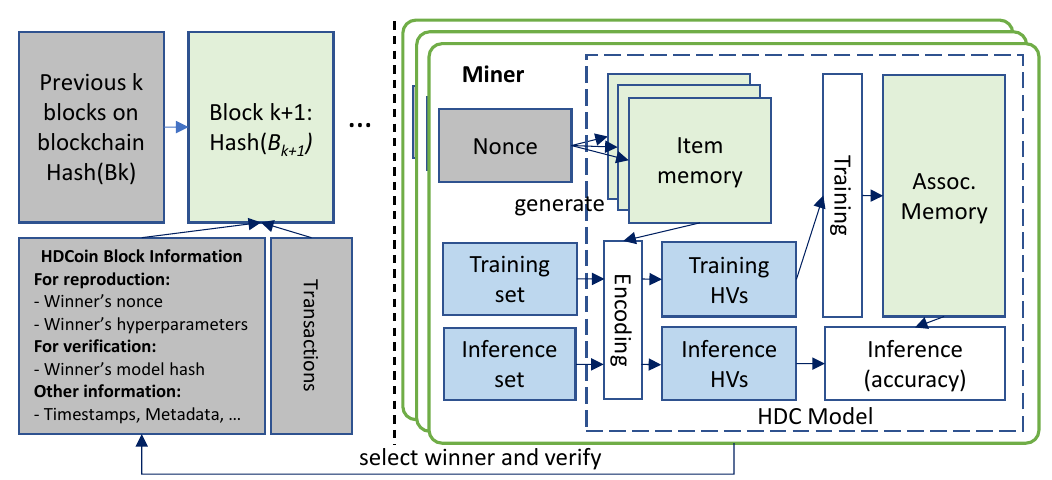}
    \caption{\model{} Framework: the blockchain (left) stores transaction data and miner verification information, while the miner (right) performs HDC model training and inference. The miner encodes data using item memory, trains on local data, and computes inference accuracy to compete for block generation. The winning model's parameters and nonce are recorded on the blockchain for reproducibility and verification.}
    \label{fig:overview}
\end{figure}

\section{\model{} Preliminaries}
\label{sec:mdev}

\subsection{HDC Notions}
\textbf{Hypervectors (HV)} are numerical vectors that are high-dimensional, holographic vectors with i.i.d. elements~\cite{kanerva2009hyperdimensional}. We can note a $d-$dimensional HV as: 

\begin{equation}
\vec{H} = \langle h_1, h_2, \dots, h_d\rangle
\end{equation}

where $h_i$ refers to the numerical elements inside the HV. In HDC, HVs are the fundamental component to represent information in different types, modalities and layers of features because of its high dimensionality.

HDC manipulates HVs using three main operations: addition, multiplication, and permutation. Addition and multiplication are element-wise operations applied between two HVs, while permutation involves cyclically shifting a single HV. These operations maintain the HV's dimensionality and support compositional encoding. To compare HVs, similarity metrics such as cosine similarity, Euclidean distance, or Hamming distance (for binary HVs) are commonly used.

\subsection{Developing HDC Model}
\label{sec:hdc_prelim}
To develop and HDC model for a classification task, there are three main steps: \textbf{Encoding}, \textbf{Training}, and \textbf{Inference}.

\textbf{Encoding} transforms an input feature vector $\vec{F}$ into its high-dimensional representation using a predefined set of operations $\Theta$ and a feature-to-HV mapping memory $\mathcal{R}$, producing an encoded HV as $\vec{H}$, which can be denoted as:

\begin{equation}
\vec{H} = \Theta(\mathcal{R}, \vec{F})
\end{equation}

\textbf{Training} involves constructing an associative memory $\mathcal{A}$ by aggregating encoded HVs by class. For a classification task with $k$ classes, the class HVs are computed by summing the HVs of all training samples belonging to each class:

\begin{equation}
\mathcal{A} = \{ \sum_{}^{}\vec{H^1}, \sum_{}^{}\vec{H^2}, \dots, \sum_{}^{}\vec{H^k} \}
\end{equation}

\textbf{Inference} encodes an unseen sample into a query HV $\vec{H_q}$ using the same $\Theta$ and $\mathcal{R}$, then compares it to all class HVs in $\mathcal{A}$ using a similarity function $\delta$. The class with the highest similarity score is assigned as the predicted label:

\begin{equation}
l = argmax(\{ \delta(\vec{H_q}, \mathcal{A})\})
\end{equation}

\textbf{Retraining} is a post-training process that updates the associative memory to further improve the performance of the HDC model. During retraining, HDC model infers using the encoded HV from training samples. If the inference is incorrect, the training sample HV will be subtracted from the incorrect class and added to the correct class using a retraining rate $\delta$. 

\begin{equation}
    \vec{H_c} = \vec{H_c} + \delta \vec{H_q}, \; \vec{H_i} = \vec{H_i} - \delta \vec{H_q}
\end{equation}

\section{\model{} Blockchain and Mining}

\begin{figure}
\centering
    \includegraphics[keepaspectratio,width=1\columnwidth]{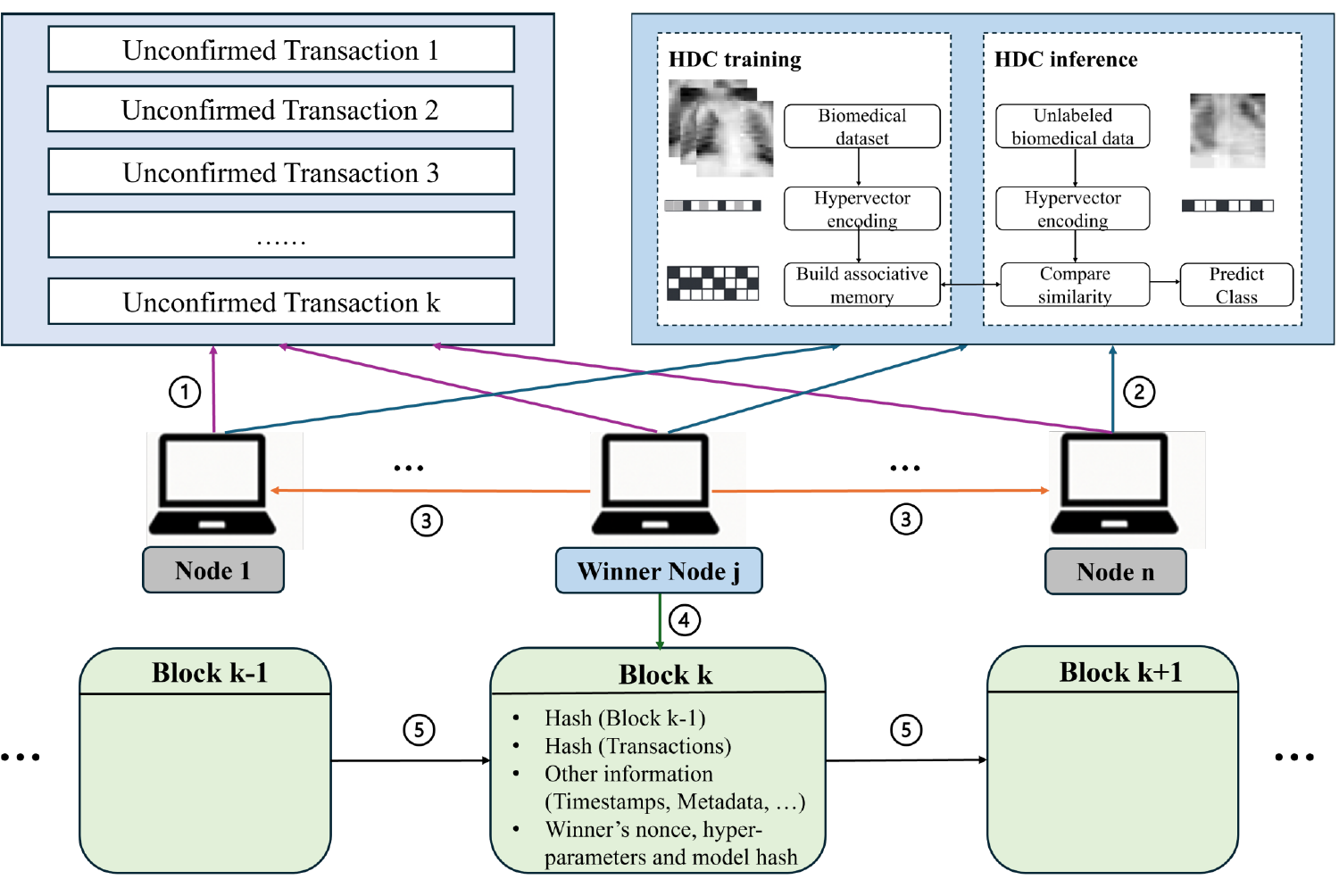}
    \caption{\model~mechanism: \ding{172} All miners validate unconfirmed transactions from transaction pool. \ding{173} All miners train HDC model on biomedical dataset and get inference accuracy. \ding{174} All miners verify the performance of the winner (train HDC model with highest accuracy). \ding{175} The winner put all the information on the blockchain. \ding{176} When one block is established, the chain moves to the next block. }
    \label{fig:chain}
\end{figure}

\textbf{Mining Cycle.} In \cref{fig:chain}, we show an overview of the main steps of a mining cycle in \model. \model ~follows the architecture of a typical blockchain network where a group of miners (nodes) compete to verify transactions. First, unconfirmed transactions are broadcast and stored in the transaction pool of the blockchain network. During the consensus process, each miner collects and validates transactions from the pool by independently performing HDC model training and inference on the biomedical dataset given the hyperparameters. Individual miners have their respective nonces to generalize their HDC model parameters such as the item memories. When mining ends, the miners  on the network will determine the winner who develops the highest accuracy HDC model and add another block to the blockchain before the next cycle starts.

In \cref{fig:overview}, we take a deeper look into each individual miner on the chain and introduce its behavior during the mining cycle. The miner first picks the transactions they wish to verify from a pool of unconfirmed transactions. At the same time, the miner obtains the dataset of a medical learning task to train an HDC model for. Once the dataset is prepared, the miner starts to randomly choose a 32-bit integer nonce and the hyperparameters (dimensions, retraining iterations, and retraining rate) of the HDC model. Using the nonce as the seed, the miner generates the item memories for the learning task. Following the steps outlined in \cref{sec:hdc_prelim}, each miner develops HDC models to compete for higher accuracy, or to verify the model from other miners.

\textbf{Trustworthiness.} When determining the winner of each cycle, every miner will attempt verify the winner by repeating the training process using the nonce from the winner to reproduce the winning model. Once a winner is verified, one additional block will be added onto the blockchain. This additional block includes the hash of the previous blocks, the hash of the selected transactions, and blockchain information such as timestamps and metadata. More importantly, the hash of the model parameters and the nonce will also be in the block so as to ensure the HDC model can be verified and reproduced at a later time by the general public.

\textbf{Adaptive Mining Difficulty.} Similar to other blockchain frameworks, \model{} supports adaptive mining difficulty. In conventional blockchains under Proof-of-Work schemes, mining difficulty is adjusted via the threshold so that blocks are found at a roughly constant rate despite changes in the total mining power of the network. In contrast, \model{} controls the mining difficulty by adjusting the minimum required dimensionality and the retraining iterations of HDC models, as higher dimensionality or more retraining iterations will require more compute overhead for training and inference. 

On the other hand, to encourage the miners to invest more compute efforts into the development of HDC models, we allow individual miners to use a higher dimensionality and more retraining iterations than the required minimum. This acts as an incentive to train models with better performance, thus will generally lead to higher winning probabilities, as we further discuss in \cref{sec:pouw_results}.

\section{Experimental Results}

\subsection{Experimental Setup}
\textbf{HDC hyperparameters.} To explore the impact of hyperparameters on \model~ mining as well as the adaptive difficulty, we sweep across a wide range of hyperparameters. For the dimensionality, we explore a range of 3000 to 15000 considering this range follows the typical dimensionality of most state-of-the-art HDC models. For retraining iterations, we explore 10 to 60 which keeps the overall mining process within a reasonable time duration. We also explore the retraining rate from 0.001 to 0.5. 

\textbf{Datasets.} We explore four biomedical datasets:
\begin{itemize}
    \item UCIHAR~\cite{anguita2013public}, a human activity recognition dataset consists of sensor readings collected from smartphones performing six different activities. There are 6213 training and 1554 testing samples.
    \item BreastMNIST~\cite{al2020dataset}, a dataset of $28\times28$ ultrasound breast images labeled as either benign or malignant for breast tumor There are 546 training and 156 testing samples.
    \item PneumoniaMNIST~\cite{kermany2018identifying}, a dataset of $28\times28$ pediatric chest X-ray images labeled as either positive or negative for pneumonia. There are 4708 training and 624 testing samples.
    \item NoduleMNIST3D~\cite{samuel2011lung}, a 3D CT scan dataset of lungs. Each scan is a 28×28×28 voxel cube labeled as either positive or negative for lung nodules. There are 1158 training and 310 testing samples.
\end{itemize}

All the experiments are conducted using a commodity laptop with Intel Core i7-12700H (2.70 GHz) CPU and 16 GB memory. Our blockchain network has 100 miners. For experimental purposes on adaptive mining difficulty, we randomly prompt some miners to opt in for higher level of configurations.

\subsection{Mining Performance}
\label{sec:dse}

\begin{figure}[htbp]
    \centering
    \includegraphics[width=0.24\textwidth]{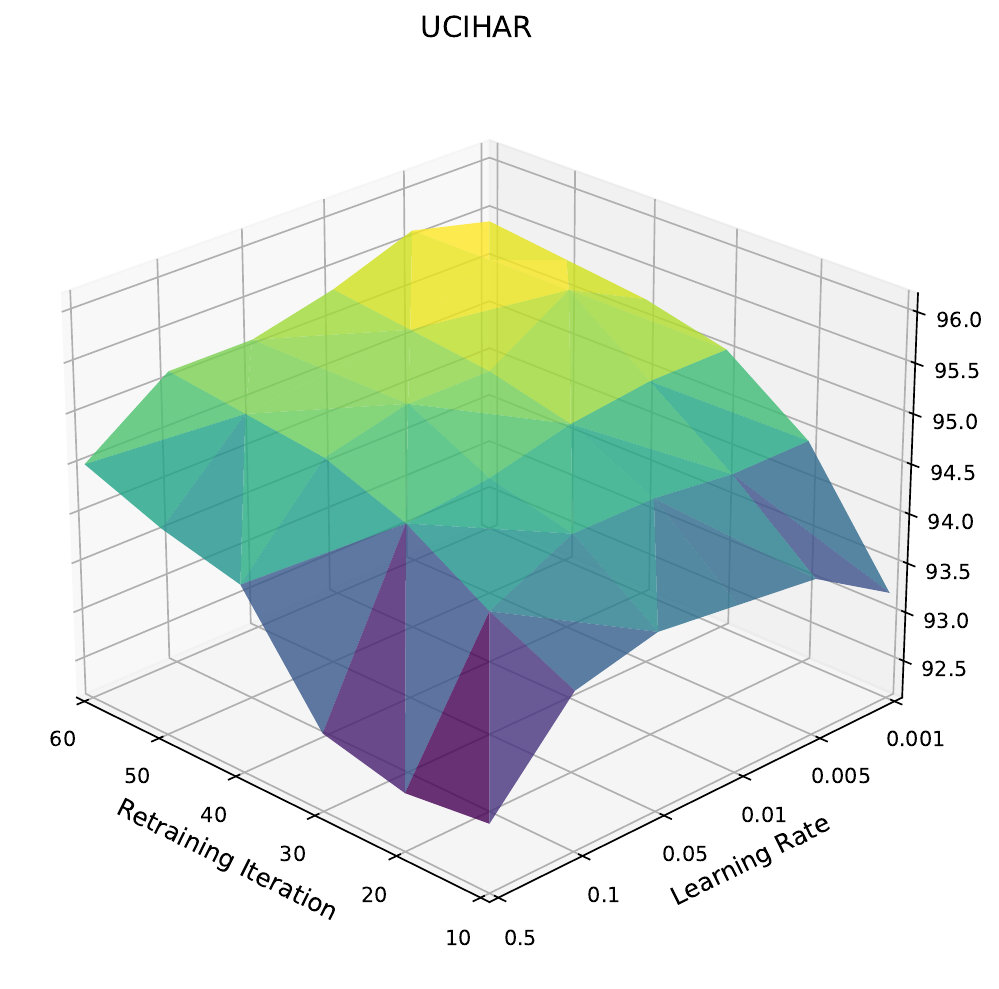}
    \includegraphics[width=0.24\textwidth]{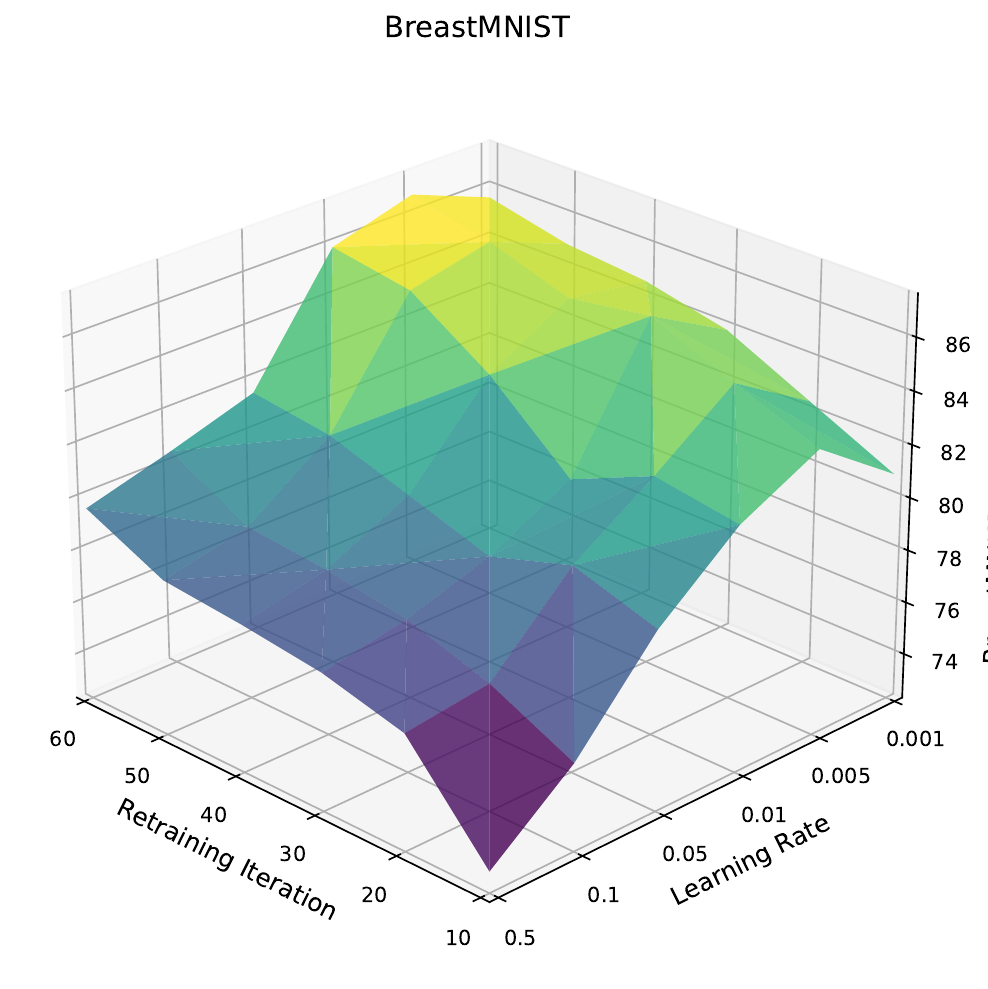}
    \includegraphics[width=0.24\textwidth]{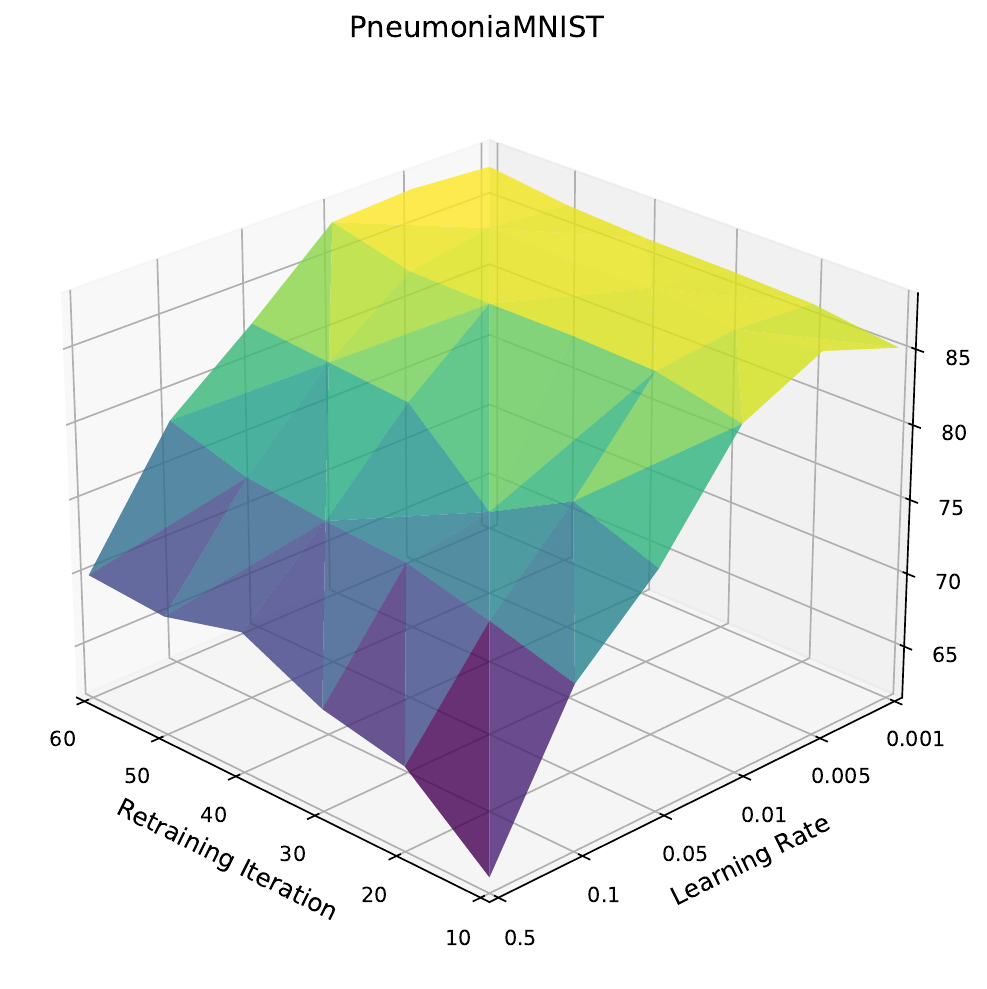}
    \includegraphics[width=0.24\textwidth]{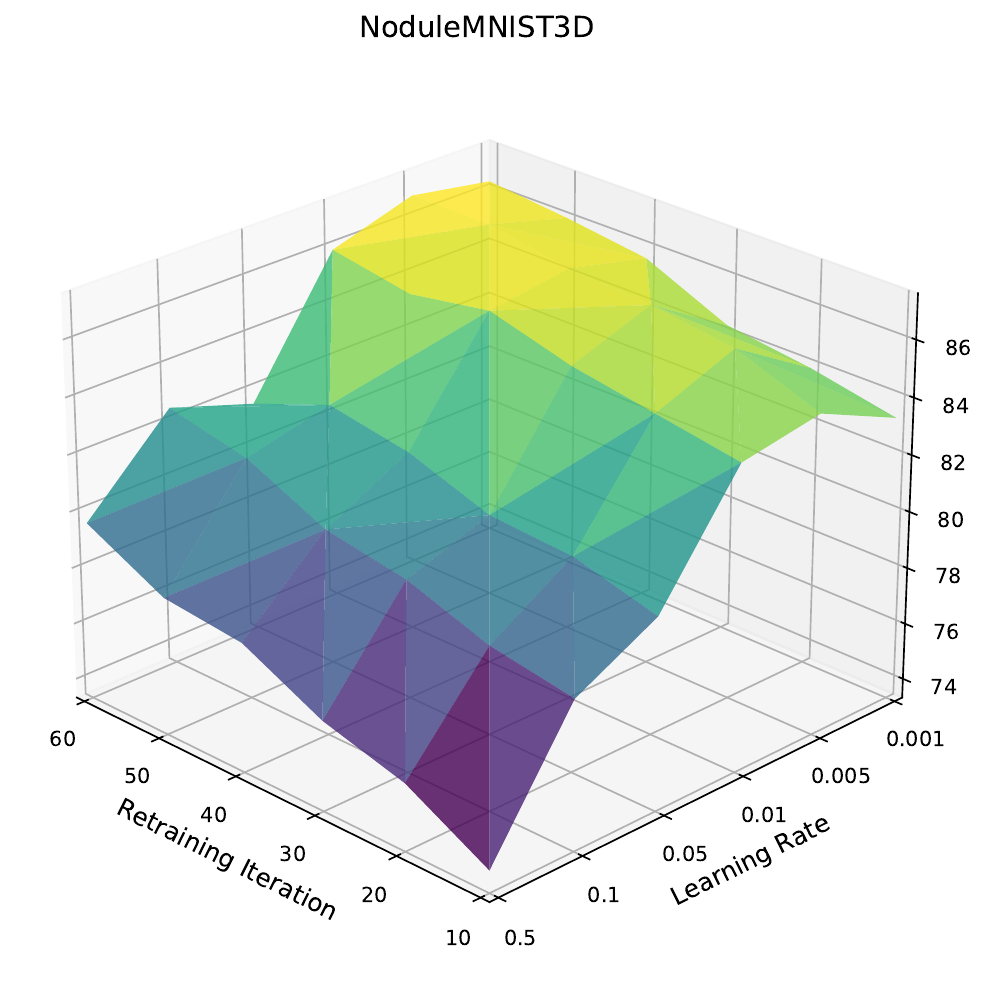}

    \caption{\model~trained HDC model accuracy under different dimensionality and retraining rates for four benchmark datasets. Generally, better model performance is observed with lower learning rates and higher numbers of retraining iterations. However, increasing hyperparameters beyond certain thresholds does not always lead to improved accuracy due to model convergence.}
    \label{fig:lr_iter_surface}
\end{figure}

In \cref{fig:lr_iter_surface}, we plot the accuracy of the HDC model trained by \model~under different hyper-parameters. We can generally observe that relatively better model performance can be achieved when using a lower learning rate and more retraining iterations. Similar to training other machine learning models, increasing those hyper-parameters will not always guarantee a better model as model capability can converge along with the progress of the model development. Therefore, we execute an empirical limit of the configurations. Together with the minimum requirements, \model~essentially provide a range of the configurations that a model can choose from, so as to train a model with acceptable performance and in the meanwhile to avoid potential model convergence.

\subsection{Adaptive Mining Difficulty}
\label{sec:adaptive}

\begin{figure}[htbp]
    \centering
    \includegraphics[width=0.23\textwidth]{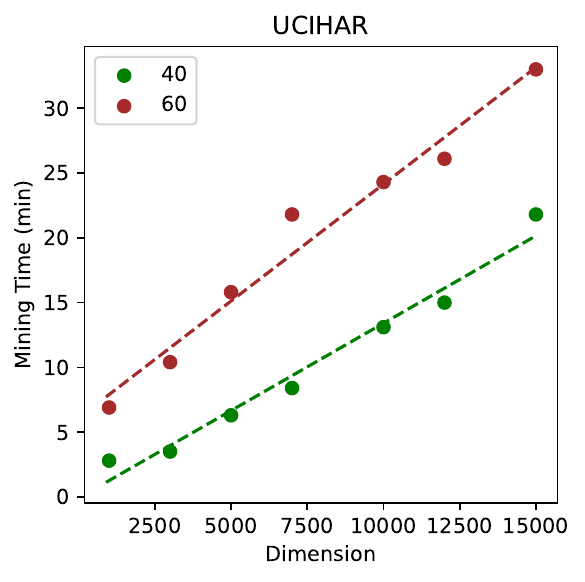}
    \includegraphics[width=0.23\textwidth]{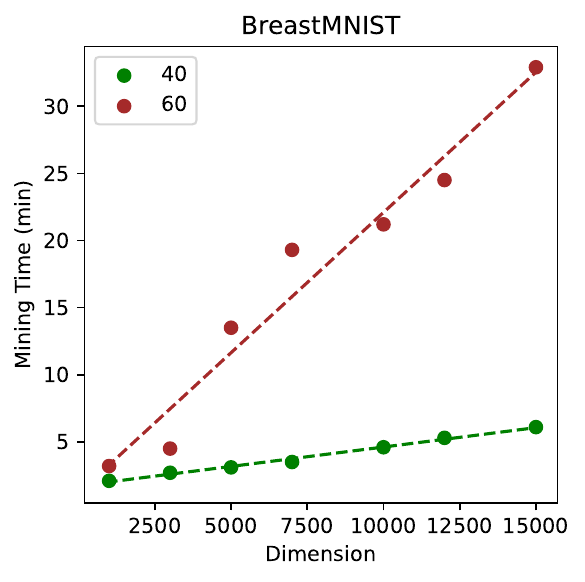}
    \includegraphics[width=0.23\textwidth]{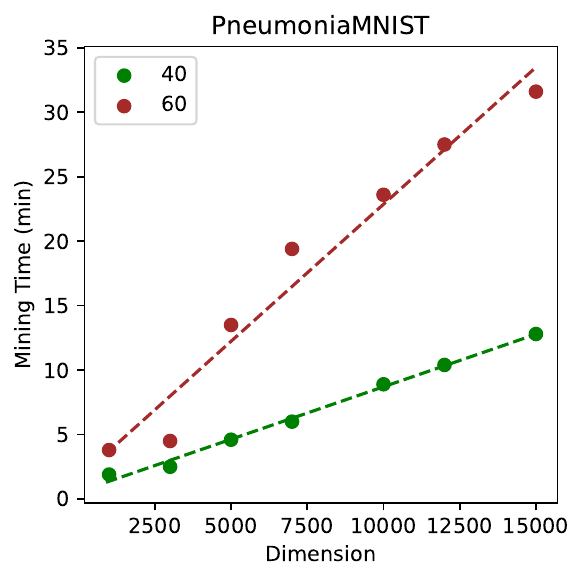}
    \includegraphics[width=0.23\textwidth]{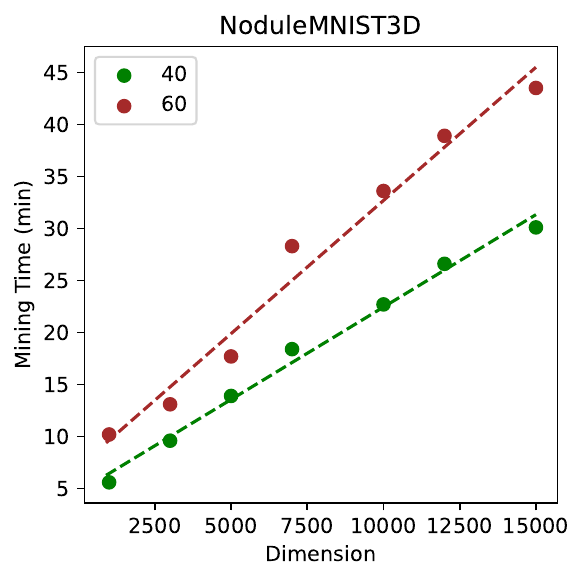}

    \caption{Adaptive Mining Difficulty in \model: increasing dimensionality (1000 - 15000) or retraining iterations (40 and 60 in this figure) leads to higher computational compute effort needed.}
    \label{fig:adaptive_mining_diff}
\end{figure}

As introduced in \cref{sec:adaptive}, mining difficulty in \model ~can be adjusted by requiring the minimum dimensionality of the HDC model and the minimum retraining iterations. In \cref{fig:adaptive_mining_diff}, we show the block time as the mining difficulty. When the dimensionality of HDC model increases, there are more elements inside HVs, thus requiring more overhead for training, inference and retraining of the model. Similarly, when \model ~requires more retraining iterations, more computations are performed within the miner during model development. Generally across the applications we evaluated, we observe that the mining difficulty linearly increases with regard to the growth of configurations.

\subsection{Fairness Proof-of-Useful-Work}
\label{sec:pouw_results}
In proof-of-work blockchains, the competition is fair in the sense that miners who invest more computational power (i.e., hash-rate) generally have a higher probability of earning rewards. Similarly in \model, as we are based on the concept of proof-of-useful-work, we encourage a similar fair competition that miners who offer more effort in training the HDC models will have a higher chance of winning. While with adaptive mining difficulty, minimum dimensionality of the HDC model and retraining iteration are required, each individual miner are allowed and encouraged to use higher dimensionality or more retraining iterations since they often (not always) lead to better performance as previously shown in \cref{fig:lr_iter_surface}.

\begin{figure}[htbp]
    \centering
    \includegraphics[width=0.9\columnwidth]{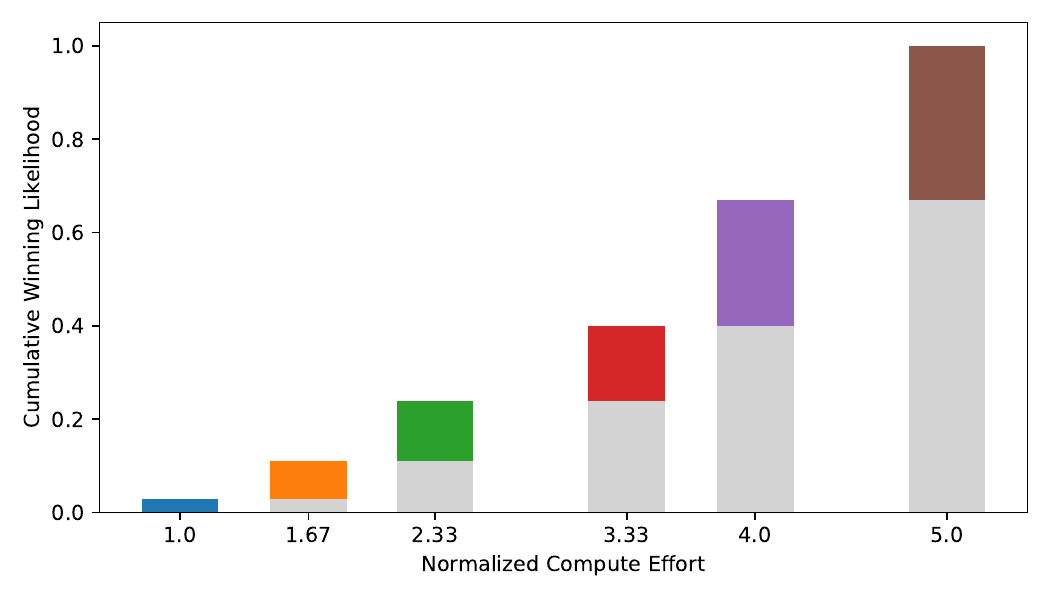}
    \caption{Cumulative winning likelihood with regard to normalized computing effort. 1.0 refers to the minimum required configuration. \model ~encourages fair competition in proof-of-useful-work: miner that invests more compute effort will have a higher likelihood of winning. }
    \label{fig:pouw_results}
\end{figure}

As illustrated in \cref{fig:pouw_results}, we provide the statistics of the probability of winning with respect to the overall compute effort of a miner. Since different applications have different architectures, dataset sizes, and HDC encoding algorithms, etc., we use the normalized compute effort to consistently describe the compute resource invested by the miner. More specifically, $1.0$ refers to the compute effort of the minimum required configurations. If the miner uses higher level of configurations, the compute effort will thus increases. Since higher configurations in general lead to better HDC models trained as discussed in \cref{fig:lr_iter_surface}, these miners will have a higher chance of winning.

\section{Conclusion}
\label{sec:cl}
In this work, we introduce \model, the first blockchain framework to integrate proof-of-useful-work with hyperdimensional computing for biomedical machine learning. By transforming mining in a blockchain into the training of HDC models, \model ~ensures trustworthy, reproducible, and fair HDC learning across distributed miners. We evaluate \model ~using four biomedical datasets and explored the HDC model performance under different configurations. We also demonstrated how adaptive mining difficulty and fair competition can be achieved through HDC-specific hyperparameters (dimensionality and retraining iterations). To the best of our knowledge, this is the first attempt to introduce blockchain into hyperdimensional computing and highlights its potential application in this emerging domain.

\bibliography{CoinHD}

\end{document}